\documentclass[useAMS,usenatbib,usegraphicx]{mn2e}

\title[ULTRACAM observations of the AXP 4U\,0142+61]{High-speed, multi-colour
optical photometry of the anomalous X-ray pulsar 4U\,0142+61 with ULTRACAM}

\author[V. S. Dhillon et al.]
{V. S. Dhillon,$^1$\thanks{E-mail: vik.dhillon@shef.ac.uk}
T. R. Marsh,$^2$ F. Hulleman,$^3$ M. H. van Kerkwijk,$^{3,4}$ A. Shearer,$^5$
\newauthor
S. P. Littlefair,$^1$ F. P. Gavriil,$^6$ V. M. Kaspi $^6$ \\\\
$^{1}$Department of Physics and Astronomy, University of Sheffield, 
Sheffield S3 7RH, UK \\
$^{2}$Department of Physics, University of Warwick, Coventry CV4 7AL, UK \\
$^{3}$Astronomical Institute, Utrecht University, PO Box 80000, 3508 TA Utrecht, The Netherlands \\
$^{4}$Department of Astronomy and Astrophysics, University of Toronto, 60 St. 
George Street, Toronto, ON M5S 3H8, Canada \\
$^{5}$Computational Astrophysics Group, Department of Information Technology, 
National University of Ireland, Galway, Ireland \\
$^{6}$Department of Physics, Rutherford Physics Building, McGill University, 
3600 University Street, Montreal, Quebec H3A 2T8, Canada \\}
\begin{document}

\date{Submitted on 2005 May 26.}

\pagerange{\pageref{firstpage}--\pageref{lastpage}} \pubyear{2005}

\maketitle

\label{firstpage}

\begin{abstract}
We present high-speed, multi-colour optical photometry of the
anomalous X-ray pulsar 4U\,0142+61, obtained with ULTRACAM on the
4.2-m William Herschel Telescope. We detect 4U\,0142+61 at magnitudes
of $i'=23.7\pm0.1$, $g'=27.2\pm0.2$ and $u'>25.8$, consistent with the
magnitudes found by \citet{hulleman04} and hence confirming their
discovery of both a spectral break in the optical and a lack of
long-term optical variability. We also confirm the discovery of
\citet{kern02b} that 4U\,0142+61 shows optical pulsations with an
identical period ($\sim8.7$\,s) to the X-ray pulsations. The rms
pulsed fraction in our data is $29\pm8$\%, 5--7 times greater than
the 0.2--8 keV X-ray rms pulsed fraction. The optical and X-ray pulse
profiles show similar morphologies and appear to be approximately in
phase with each other, the former lagging the latter by only
$0.04\pm0.02$ cycles. In conjunction with the constraints imposed by
X-ray observations, the results presented here favour a magnetar
interpretation for the anomalous X-ray pulsars.
\end{abstract}


\section{Introduction}

More than 100 X-ray pulsars are currently known. The vast majority of
these are found in low-mass and high-mass X-ray binaries (LMXBs and
HMXBs), and are hence powered by accretion onto a rotating, magnetised
neutron star. There exists a small group of 8 X-ray pulsars, however,
that exhibit properties very much at variance with those of the
accreting pulsars in X-ray binaries. These so-called Anomalous X-ray
Pulsars (AXPs) all have $\sim 5-12$\,s spin periods which decrease
steadily with time, soft (and relatively low-luminosity) X-ray
spectra, no radio emission, and tend to be associated with supernova
remnants in the galactic plane. Most importantly, the AXPs show no
evidence of a binary companion. For a recent review of AXPs, see 
\citet{woods04}.

This latter fact prompted a variety of models based on isolated
neutron stars and white dwarfs (see the review by \citealt{israel02}),
but these run into difficulty on energetic grounds: the loss of
rotational energy, which powers radio pulsars like the Crab, is orders
of magnitude too small to power the observed X-ray luminosity of the
AXPs. An additional energy source is therefore required, for which two
competing models seem to have emerged: accretion from a fossil disc or
ultra-strong magnetic fields. In the former scenario, an isolated
neutron star accretes from a fossil disc, such as might be produced
through fall-back of material after a supernova explosion or left over
from a common-envelope phase which destroyed the companion star. In
the latter scenario, AXPs are ``magnetars'', isolated neutron stars
with enormous ($B\sim10^{14}-10^{15}$ G) magnetic fields. It is the
decay of the magnetic field which heats the neutron star surface,
causing it to emit thermal radiation in the X-rays. Non-thermal
emission is then produced by particles accelerated in the
magnetosphere by the Alfv\'{e}n waves from small-scale fractures on
the neutron star surface \citep{thompson96} or Comptonization of
thermal photons by magnetospheric currents \citep{thompson02}.

The magnetar model has begun to dominate the literature in recent
years. There are sound theoretical reasons for why this is so, as the
magnetar model now appears to be able to explain the observational
properties of several categories of supposedly young neutron stars
that are not powered by rotation, including the AXPs and the Soft
Gamma Repeaters (SGRs). Such unification is supported by the recent
discovery of SGR-like bursts in AXPs (\citealt{gavriil02a};
\citealt{kaspi03}), which suggests there might be an evolutionary link
between AXPs and SGRs (see \citealt{mereghetti02} and references
therein).

But what other observational evidence is there to support the magnetar
model of AXPs? Belief in the magnetar model rests partly on the
failure of the accretion model to explain the faintness of the
optical/infrared counterparts, which sets strong limits on the size of
an accretion disc (e.g. \citealt{hulleman00}), and the fact that the
pulsed fraction of optical light is significantly greater than it is
in X-rays, ruling out reprocessing of X-rays in a disc as its origin
(\citealt{kern02b}; but see \citealt{ertan04}). Both of these optical
constraints have been obtained via observations of the brightest known
AXP, 4U\,0142+61. In this paper we report on new high-speed,
multi-colour optical observations of this object, obtained with the
aim of confirming the high optical pulsed fraction observed by
\citet{kern02b}.

\section{Observations and data reduction}
\label{obsred}

\begin{figure*}
\centering
\includegraphics[width=10cm,angle=270]{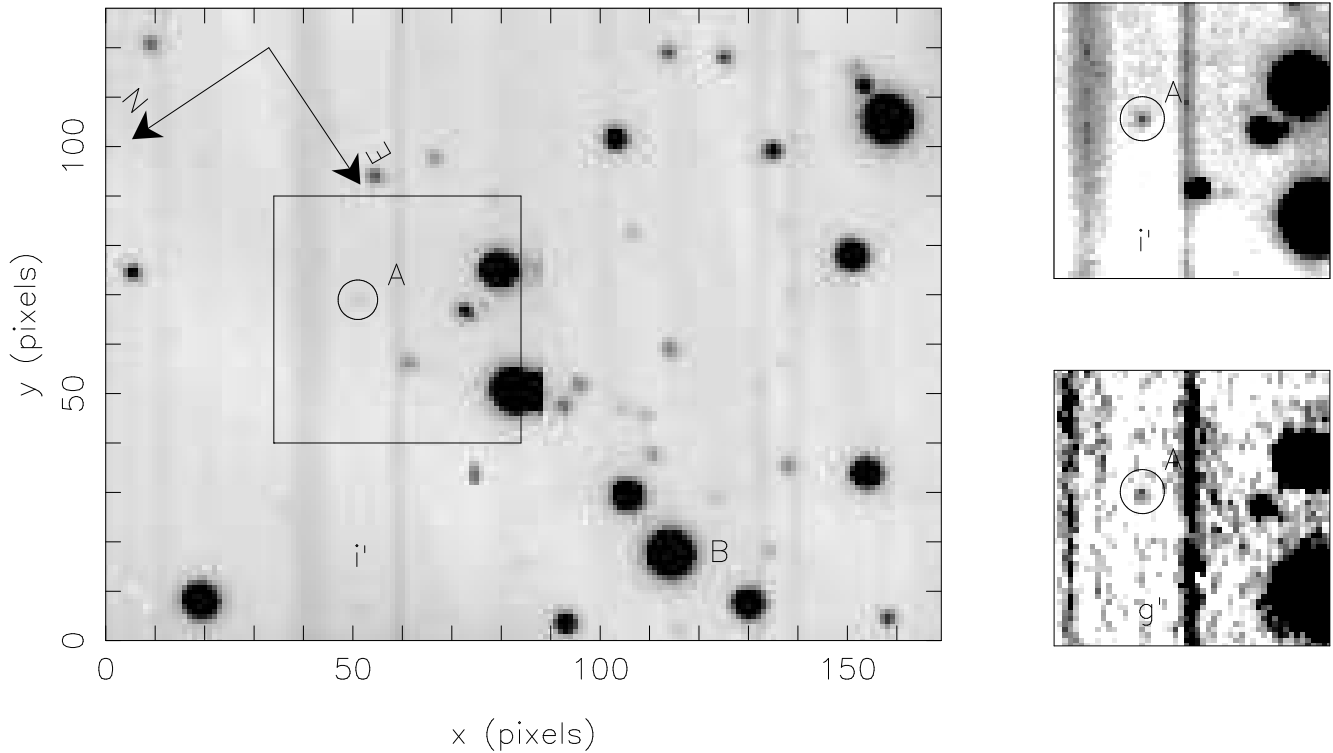}
\caption{Left: Summed $i'$ image from the night of 12/09/02, with a
total exposure time of 15046~s ($\sim 4$~h). Star A is 4U\,0142+61
and star B is the comparison/reference star (see \citet{hulleman04}
for coordinates and magnitudes). The orientation arrows represent 10
arcseconds on the sky.  For clarity, only a portion of the two
ULTRACAM windows is shown. Note that there is no gap between the
windows, but a faint discontinuity between them can be seen running
down the centre of the image, due to the fact that each window is read
out via a separate channel. Right: Higher contrast plots of the field
around 4U\,0142+61 (star A), showing the summed $i'$ (top) and $g'$
images (bottom) from the night of 12/09/02. The box in the left-hand
image shows the portion of the field shown (at the same scale) in the
right-hand images.}
\label{fig1}
\end{figure*}

The observations of 4U\,0142+61 presented in this paper were obtained
with ULTRACAM (\citealt{dhillon01}, \citealt{beard02}) at the
Cassegrain focus of the 4.2-m William Herschel Telescope (WHT) on La
Palma.  ULTRACAM is a CCD camera designed to provide imaging
photometry at high temporal resolution in three different colours
simultaneously. The instrument provides a 5 arcminute field on its
three $1024\times1024$ E2V 47-20 CCDs (i.e. 0.3
arcseconds/pixel). Incident light is first collimated and then split
into three different beams using a pair of dichroic beamsplitters. For
the observations presented here, one beam was dedicated to the SDSS
$u'$ (3543\AA) filter, another to the SDSS $g'$ (4770\AA) filter and
the third to the SDSS (7625\AA) $i'$ filter. Because ULTRACAM employs
frame-transfer chips, the dead-time between exposures is negligible:
we used ULTRACAM in its two-windowed mode, each of $100\times200$
pixels, resulting in an exposure time of 0.48~s and a dead-time of
0.025~s. A total of 30\,618 and 31\,304 frames of 4U\,0142+61 were
obtained on the nights of 2002 September 10 and 12, respectively, with
each frame time-stamped to a relative accuracy of better than
50~$\mu$s using a dedicated GPS system.\footnote{The absolute timing
accuracy of ULTRACAM was verified with contemporaneous observations of
the Crab pulsar. Our observed time of optical pulse maximum was found
to agree with the ephemeris of \citet{lyne05} to better than 1
millisecond (the quoted error in the Crab pulsar ephemeris during
September 2002).} Both sets of data were obtained in photometric
conditions, with no moon and $i'$-band seeing of 0.75 and 0.65
arcseconds on 10/09/02 and 12/09/02, respectively.

A portion of the summed $i'$ image from the night of 12/09/02 is shown
in figure~\ref{fig1}. The vertical streaks are due to light from
bright stars falling on the active area of the chip above the CCD
windows. The data we obtained on 10/09/02 (not shown in
figure~\ref{fig1}) suffer from streaks passing through the position of
4U\,0142+61, increasing the background noise level significantly. As a
result, we rotated the Cassegrain rotator in advance of our
observations on 12/09/02 so that no streaks passed through
4U\,0142+61. For this reason, the data obtained on 12/09/02 are of a
much higher quality than the data obtained on 10/09/02. Note that the
vertical streaking problem has since been rectified in ULTRACAM by
provision of an adjustable focal-plane mask which blocks the light
from bright stars (and the sky) above the CCD windows
\citep{stevenson04}.

The data were reduced using the ULTRACAM pipeline software. All frames
were first debiased and then flat-fielded, the latter using the median
of twilight sky frames taken with the telescope spiralling. We then 
extracted light curves of 4U\,0142+61 using two different techniques:

\subsection{Technique (i)}
\label{tech1}

\citet{kern02b} obtained their light curve of 4U\,0142+61 by
synchronising the CCD clocks in their camera to the X-ray spin
period of 4U\,0142+61, resulting in the accumulation of 10 on-chip
phase bins.  This has the advantage of reducing detector noise, but
the potential disadvantage that a period must be assumed before the
data have been taken and, if the period is wrong, the true pulse
profile is unrecoverable.

\begin{table}
\centering
\caption{Updated ephemeris for 4U\,0142+61 spanning the optical
observations described in section~\ref{obsred}, based on the
monitoring campaign described in \citet{gavriil02b}. BMJD refers
to the Barycentric-corrected Modified Julian Date on the
Barycentric Dynamical Timescale (TDB). TOA refers to the pulse time 
of arrival (see \citet{gavriil02b} for details). The errors on the
last two digits of each parameter are given in parentheses.}
\begin{tabular}{lr}
& \\
\hline
\hspace*{4cm} & \\
BMJD range\dotfill & $51\,610.636-53\,401.184$ \\
TOA arrival points\dotfill & 79 \\
$\nu$ (Hz)\dotfill & 0.11509507445(18) \\
$\dot\nu$ ($10^{-14}$ Hz s$^{-1}$)\dotfill & --2.66478(36) \\
$\ddot\nu$ ($10^{-24}$ Hz s$^{-2}$)\dotfill & 5.08(23) \\
Epoch (BMJD)\dotfill & 52\,506.9748874274228 \\
rms residual (cycles)\dotfill & 0.031 \\
& \\
\hline
\end{tabular}
\end{table}

To mimic the \citet{kern02b} technique, we assumed a spin period for
4U\,0142+61 on 12/09/02 of 8.688473130~s, which was calculated from
the updated X-ray ephemeris given in table~1. Note that this ephemeris
spans our WHT observations and is hence more reliable for our purposes
than using the ephemeris of \citet{gavriil02b} adopted by
\citet{kern02b}. Each ULTRACAM data frame was then added to one of 10
evenly-spaced phase bins covering the spin cycle of 4U\,0142+61,
resulting in 10 high signal-to-noise data frames. An optimal
photometry algorithm \citep{naylor98} was then used to extract the
counts from 4U\,0142+61 and a bright comparison star 24 arcseconds to
the east of the AXP (see figure~\ref{fig1}), the latter acting as the
reference for the profile fits and transparency-variation
correction. The position of 4U\,0142+61 relative to the comparison
star was determined from a sum of all the images, and this offset was
then held fixed during the reduction so as to avoid aperture
centroiding problems. The sky level was determined from a clipped mean
of the counts in an annulus surrounding the target stars, and
subtracted from the object counts.

\subsection{Technique (ii)}
\label{tech2}

The second approach we took to light curve extraction was
identical to that described above, except that we omitted the
phase-binning step and simply performed optimal photometry on the
61922 individual ULTRACAM data frames. In other words, we made no
assumption about the spin period of 4U\,0142+61.

\section{Results}

\subsection{Magnitudes}
\label{mags}

We were unable to detect 4U\,0142+61 in $u'$, at a detection limit of
$u'>25.8$. We did, however, clearly detect it in $g'$ and $i'$ on
both nights at magnitudes of $g'=27.2\pm0.2$ and $i'=23.7\pm0.1$, as
shown in figure~\ref{fig1}. \citet{hulleman04} measured $g' \sim 26.9$
and $i' \sim 23.7$ (where we have converted their $BVRI$
Johnson-Morgan-Cousins magnitudes to SDSS magnitudes using the
transformation equations of \citealt{smith02}), indicating that
4U\,0142+61 was approximately the same magnitude during our
observations.

\subsection{Pulse profiles}

The two data reduction techniques described in section~\ref{obsred}
result in two different pulse profiles for 4U\,0142+61.

\subsubsection{Technique (i)}
\label{techi}

\begin{figure}
\centering
\includegraphics[width=6.4cm,angle=270]{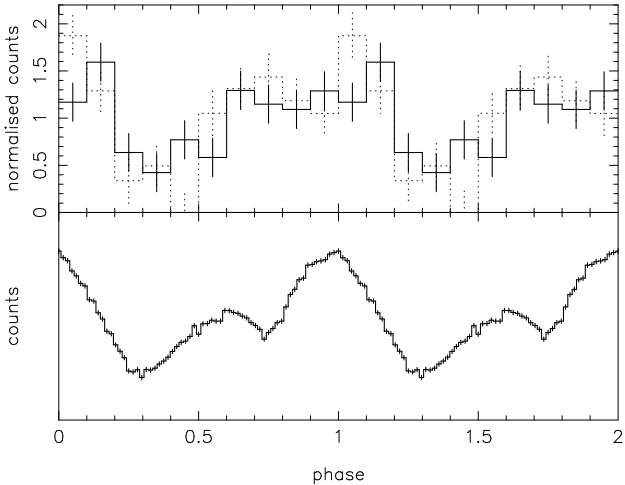}
\caption{Top: Pulse profiles of 4U\,0142+61 in the $i'$-band on
12/09/02 (solid line), obtained using technique (i)
(section~\protect{\ref{tech1}}). The dotted line shows the poorer
quality light curve we obtained on the night of 10/09/02,
demonstrating the repeatability of the pulse profile. Each pulse
profile was first corrected for transparency variations using the
comparison star (star B in figure~\ref{fig1}), although the correction
made only a negligible difference to the light curves.  The pulse
profiles were then normalised by dividing by the mean number of
counts.  Note that the formal error bars on these pulse profiles were
unreliable (most probably due to the vertical streaks shown in
figure~\ref{fig1}), and hence the error bars shown have been
calculated from the scatter in the light curve extracted using
technique (ii) (section~\protect{\ref{tech2}}). Bottom: Averaged X-ray
pulse profile of 4U\,0142+61 in the 2--10 keV energy band, which is an
updated version of the profile presented in \citet{gavriil02b}. Note
that it is not possible to estimate the X-ray pulsed fraction from
this profile as the background level (i.e. the minimum flux) in the
X-ray pulse profile is unrelated to the pulsar -- see
section~\ref{disc} for details). For this reason, no scale is given on
the ordinate.}
\label{fig2}
\end{figure}

The first technique produced the pulse profiles shown in the top panel
of figure~\ref{fig2}. As expected, the light curve of 12/09/02 is of a
significantly higher quality than that of 10/09/02, but both show
approximately the same morphology as the optical pulse profile
presented by \citet{kern02b}, exhibiting a broad (arguably
double-humped) structure with peaks around phases 0.65 and 1.15 and a
minimum around phase 0.35. These phases are different to the
corresponding phases in the pulse profile of \citet{kern02b}, but this
is to be expected given that, as discussed by \citet{kern02b}, their
optical observations were obtained outside the span of the ephemeris
they used and the source does exhibit some timing noise
\citep{gavriil02b}.  Our timing solution, on the other hand, is based
on the updated ephemeris for 4U\,0142+61 presented in table~1, which
spans our WHT observations and is hence reliable.

There is some similarity in the morphologies of the optical pulse
profile shown in the top panel of figure~\ref{fig2} and the 2--10 keV
X-ray pulse profile shown below it, where the latter is an updated
version of the data presented by \citet{gavriil02b}. Both profiles
share a similar broad/double-humped morphology. Moreover, since the
X-ray light curve shown in figure~\ref{fig2} has also been phased
using the ephemeris given in table~1, it can be seen that the optical
and X-ray pulse profiles are approximately in phase with each
other. To quantify this, the optical pulse profile was
cross-correlated with the X-ray pulse profile. The resulting peak in
the cross-correlation function was fitted with a parabola to derive a
shift of $0.04\pm0.02$ cycles (i.e. $0.35\pm0.17$\,s), where a
positive phase shift implies that the optical pulse profile lags the
X-ray pulse profile. This result is only marginally significant (at
the 2$\sigma$ level), due to the low signal-to-noise and time
resolution of the optical data, and additional data will be required
in order to confirm that the phase shift is significantly different
from zero (discounting the unlikely situation in which the time delay
is approximately equal to some multiple of the spin period).

The modulation amplitude of the pulses presented in figure~\ref{fig2}
can be measured using a peak-to-trough pulsed fraction, $h_{pt}$,
defined as follows:

\begin{equation}
h_{pt}=\frac{F_{max}-F_{min}}{F_{max}+F_{min}},
\end{equation}

\noindent where $F_{max}$ and $F_{min}$ are the maximum and minimum
flux in the pulse profile, respectively. We find a value of
$h_{pt}=58\pm16$\% on 12/09/02, higher than the pulsed fraction of
$h_{pt}=27\pm8\%$ derived by \citet{kern02b}, although the difference
between the two values is only marginally significant ($31\pm18\%$,
i.e. $<2\sigma$).  There are a number of factors which might
contribute to a higher optical pulsed fraction in our data:

\begin{itemize}
\item The pulsed fraction measured from our data refers to the $i'$
band, whereas that of \citet{kern02b} is for white light
(4000--10000\AA). If the optical pulsed fraction varies with
wavelength, this could be the source of the discrepancy. Note that our
$g'$ data were too faint to extract a pulse profile from,
unfortunately, so we are not in a position to test this
explanation.
\item Even a small error in the assumed period on which the data is
phase-binned can result in a smearing of the pulse profile and hence a
reduction in the measured pulsed fraction. We have simulated this
effect and find that to reduce our pulsed fraction to the level
observed by \citet{kern02b}, the period must be in error by greater
than $\sim 0.003$~s. This is three orders of magnitude greater than the
timing accuracy achieved by the instrumentation used by
\citet{kern02b} and hence an error in the period used to phase bin the
data is an unlikely source of the discrepant pulsed fractions.
\item The higher pulsed fraction in our data might be due either to a
decrease in the unpulsed optical component or an increase in the
pulsed optical component. \citet{hulleman04} found no evidence for
long-term $R$-band variability in 4U\,0142+61, down to a 2-$\sigma$
limit of 0.09 magnitudes, and our magnitude estimates
(section~\ref{mags}) appear to support this conclusion.  Note,
however, that \citet{hulleman04} did find long-term variability of
$\sim 0.5$ magnitude in their $K$-band observations of 4U\,0142+61,
which they tentatively attributed to the occurence of SGR-like bursts
in 4U\,0142+61. X-ray observations of this source have not shown such
bursts, but this might be due to their (expected) low amplitude and
the efficiency of the X-ray monitoring.
\item The peak-to-trough pulsed fraction defined in equation~1
effectively adds any noise present in the light curve to the true
pulsed fraction, thereby tending to increase the resulting
measurement. A more robust estimate is given by the root-mean-square
(rms) pulsed fraction, $h_{rms}$, defined as follows:

\begin{equation}
h_{rms}= \frac{1}{\bar{y}} \left[ \frac{1}{n} \sum_{i=1}^{n}(y_i-\bar{y})^2-\sigma_i^2 \right]^{\!\frac{1}{2}},
\end{equation}

\noindent where $n$ is the number of phase bins per cycle, $y_i$ is
the number of counts in the $i^{\rm th}$ phase bin, $\sigma_i$ is the
error on $y_i$ and $\bar{y}$ is the mean number of counts in the
cycle. As expected, measuring the optical pulsed fraction in this way
gives a lower value of $h_{rms} = 29\pm8$\%. This is much closer to
the value derived by \citet{kern02b}, but it should be stressed that
these authors measured the peak-to-trough pulsed fraction (equation
1), not the rms pulsed fraction (equation 2).
\item The higher pulsed fraction in our data could be due to some
systematic problem with the sky subtraction. We consider this to be
unlikely, however, as one would not then expect our magnitude
estimates to agree with those of \citet{hulleman04}.
\end{itemize}

\subsubsection{Technique (ii)}

\begin{figure}
\centering
\includegraphics[width=6.3cm,angle=270]{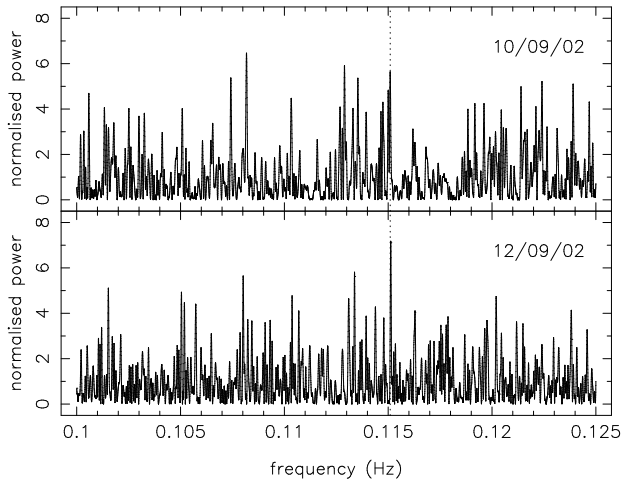}
\caption{Lomb-Scargle periodograms of 4U\,0142+61 in the $i'$-band on
10/09/02 (top panel) and 12/09/02 (bottom panel), obtained using the
light curves from technique (ii) (section~\protect{\ref{tech2}}).
The dotted line shows the predicted X-ray pulse frequency of
0.11509502130~Hz on 12/09/02, calculated from the ephemeris given
in table~1.}
\label{fig3}
\end{figure}

The second data reduction technique (section~\ref{tech2}) can be used
to provide a check on the reliability of the optical pulse profile
shown in figure~\ref{fig2}. To do this, it is first necessary to fold
the extracted light curve on the pulse period. Rather than do this by
adopting the X-ray ephemeris given in table~1, as we did in
figure~\ref{fig2}, we can instead determine the pulse period directly
from our optical data using a periodogram and then fold the data on
this period.

Figure~\ref{fig3} shows the Lomb-Scargle periodograms \citep{press89}
for the 30\,618 and 31\,304 points in the $i'$ light curves obtained on
10/09/02 and 12/09/02, respectively. The $g'$ light curves were
unfortunately too noisy to perform such an analysis. The light curves
were first corrected for transparency variations and then detrended by
subtracting their mean level. The highest peak in the resulting
periodogram of 12/09/02 occurs at a period of $8.687\pm0.002$~s, where
the error is given by the width ($\sigma$) of a Gaussian fit to the
peak in the periodogram. This period is consistent with the X-ray
pulse period given in table~1. Although noisier, an
equivalent peak is also present in the periodogram of 10/09/02, with a
period of $8.688\pm0.002$~s, thereby confirming that we have indeed
detected the X-ray pulsation of 4U\,0142+61 in the optical. We further
tested the robustness of our period detection by constructing 10000
randomised light curves from the original light curves by randomly
re-ordering the $y$-axis points. Only 0.12\% of the resulting 10000
periodograms for the 12/09/02 dataset showed a higher peak at 8.687~s,
and only 0.38\% showed a higher peak at 8.688~s in the 10/09/02
dataset.

\begin{figure}
\centering
\includegraphics[width=5.6cm,angle=270]{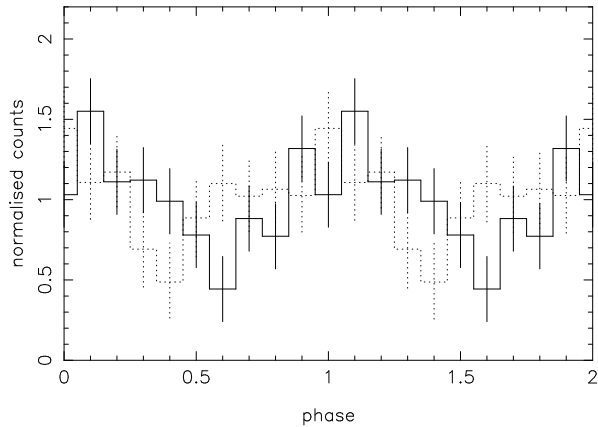}
\caption{Pulse profiles of 4U\,0142+61 in the $i'$-band on 12/09/02,
obtained using technique (ii) (section~\protect{\ref{tech2}}) and
folding the resulting data on the optically-determined period of
$8.687$~s (solid line) and on the X-ray period of 8.688473130~s
(dotted line). The data were first corrected for transparency
variations using the comparison star (star B in
figure~\ref{fig1}). The pulse profiles were then normalised by
dividing by the mean number of counts.  Note that the formal error
bars on these pulse profiles were unreliable (most probably due to the
vertical streaks shown in figure~\ref{fig1}), and hence the error bars
shown have been calculated from the scatter in the unfolded light
curve.}
\label{fig4}
\end{figure}

Folding the $i'$ light curve of 12/09/02 on the derived optical pulse
period of 8.687~s gives the pulse profile shown in
figure~\ref{fig4}. The same data folded on the X-ray pulse period of
8.688473130~s is shown for comparison. Note that the phasing of both
profiles can be directly compared to that in figure~\ref{fig4}, as
all of the data were folded using the zero point given in table~1.

As one would expect, the $i'$-data folded on the X-ray pulse period
(dotted line in figure~\ref{fig4}) is in excellent agreement with that
presented in figure~\ref{fig2}, in terms of morphology, phasing and
pulsed fraction ($h_{pt}=50\pm20$\%). The $i'$-data folded on the
optically-determined pulse period (solid line in figure~\ref{fig4})
shares approximately the same phase of pulse maximum and pulsed
fraction ($h_{pt}=56\pm16$\%), but the morphology
is slightly different. In particular, the shape and phase of pulse
minimum is very different, This is to be expected, however, given that
the data have been folded on the optically-derived period of 8.687~s,
which is much less accurate than the X-ray period due to the lower
quality of the optical data.

\section{Discussion and conclusions}
\label{disc}

Using two different data reduction techniques we have shown that the
optical light from 4U\,0142+61 pulsates on the X-ray period, thereby
confirming the discovery of \citet{kern02b}. The morphologies of the
2--10 keV and $i'$ pulse profiles are quite similar, both exhibiting a
broad/double-humped structure. The optical lags the X-rays by only
$0.04\pm0.02$ cycles ($0.35\pm0.17$\,s), i.e. there is no strong
evidence for a phase shift between the two pulse profiles. The most
reliable value we have derived for the optical pulsed fraction is
$h_{rms}=29\pm8$\%, as this is the more robust rms figure (as opposed
to the peak-to-trough value) and has been obtained by folding the best
dataset, that of 12/09/02, on the accurately known X-ray period (see
figure~\ref{fig1}). The X-ray pulsed fraction of 4U\,0142+61 cannot be
measured from the X-ray pulse profile shown in figure~\ref{fig2},
unfortunately, as the background level (i.e. the minimum flux) in
these data is unrelated to the pulsar. This is because the X-ray data
were obtained with the Proportional Counter Array (PCA) on the Rossi
X-ray Timing Explorer (RXTE), which has approximately a 1$^{\circ}$
field of view and no imaging capability. Instead, we turn to the work
of \citet{patel03}, who reported X-ray pulsed fractions for
4U\,0142+61 of $h_{rms}=4.6\pm0.5$\% between 0.2--1.3 keV,
$h_{rms}=4.1\pm0.4$\% between 1.3--3.0 keV and $h_{rms}=5.6\pm1.0$\%
between 3.0--8.0 keV; \citet{patel03} also quote corresponding
peak-to-trough pulsed fractions of $h_{pt}=8.4\pm1.6$\%,
$h_{pt}=7.4\pm1.2$\% and $h_{pt}=11.7\pm3.2$\%. These values
demonstrate that the optical rms pulsed fraction we have measured is
5--7 times greater than the X-ray rms pulsed fractions, consistent
with the factor of 5--10 times derived by \citet{kern02b} from their
peak-to-trough pulsed fraction measurements.

We have measured $g'$ and $i'$ magnitudes consistent with the $BVRI$
magnitudes found by \citet{hulleman04}, supporting their finding that,
although variable in the infrared and X-rays, 4U\,0142+61 does not
appear to show long-term variability in the optical part of the
spectrum.  Moreover, the fact that we have used the $B$ and $V$
magnitudes of \citet{hulleman04} to derive a $g'$ magnitude consistent
with our own confirms that the spectral break between $B$ and $V$
found by \citet{hulleman04} is real.

The optical observations presented in this paper therefore lend
additional weight to the arguments given by \citet{hulleman00},
\citet{kern02b} and \citet{hulleman04} that the AXP's are best
explained by the magnetar model, mainly thanks to the failure of most
alternative models to explain the observations (a notable exception is
the disc-star dynamo gap model of \citealt{ertan04}). In particular,
fall-back accretion disc models (e.g. \citealt{perna00}) fail because
the optical flux is assumed to be due to reprocessing of the X-ray
flux in the disc and therefore would not be expected to show either an
optical pulsed fraction significantly in excess of the X-ray pulsed
fraction (see \citealt{kern02b}) or a non-thermal spectral energy
distribution in the optical. In addition, such reprocessing might also
be expected to result in the optical pulses lagging the X-ray pulses
in phase by an amount depending on the light-travel time to the
reprocessing structure, the reprocessing timescale within it and its
location with respect to the X-ray source and Earth.  We have shown
that there is no strong evidence for a phase shift between the optical
and X-ray pulses, lending further evidence in support of the magnetar
model.

\section*{Acknowledgments}

We thank the referee for his valuable comments on the original
manuscript.  TRM acknowledges the support of a PPARC Senior Research
Fellowship.  SPL is supported by PPARC grant
PPA/G/S/2003/00058. ULTRACAM is supported by PPARC grant
PPA/G/S/2002/00092.  The William Herschel Telescope is operated on the
island of La Palma by the Isaac Newton Group in the Spanish
Observatorio del Roque de los Muchachos of the Instituto de
Astrof\'{i}sica de Canarias.

\bibliographystyle{mn2e}
\bibliography{abbrev,refs}

\begin{thebibliography}{}

\bibitem[\protect\citeauthoryear{Beard, Vick, Atkinson, Dhillon, Marsh, McLay,
  Stevenson \& Tierney}{Beard et~al.}{2002}]{beard02}
Beard S.~M.,  Vick A.~J.~A.,  Atkinson D.,  Dhillon V.~S.,  Marsh T.~R.,  McLay
  S.,  Stevenson M.~J.,    Tierney C.,  2002, in Lewis H.,  ed., Advanced
  Telescope and Instrumentation Control Software II. SPIE, 4848, p.~218

\bibitem[\protect\citeauthoryear{{Dhillon} \& {Marsh}}{{Dhillon} \&
  {Marsh}}{2001}]{dhillon01}
{Dhillon} V.,  {Marsh} T.,  2001, New Astronomy Review, 45, 91

\bibitem[\protect\citeauthoryear{Ertan \& Cheng}{Ertan \&
  Cheng}{2004}]{ertan04}
Ertan {\" U}.,  Cheng K.~S.,  2004, ApJ, 605, 840

\bibitem[\protect\citeauthoryear{Gavriil \& Kaspi}{Gavriil \&
  Kaspi}{2002}]{gavriil02b}
Gavriil F.~P.,  Kaspi V.~M.,  2002, ApJ, 567, 1067

\bibitem[\protect\citeauthoryear{Gavriil, Kaspi \& Woods}{Gavriil
  et~al.}{2002}]{gavriil02a}
Gavriil F.~P.,  Kaspi V.~M.,    Woods P.~M.,  2002, Nat, 419, 142

\bibitem[\protect\citeauthoryear{Hulleman, van Kerkwijk \& Kulkarni}{Hulleman
  et~al.}{2000}]{hulleman00}
Hulleman F.,  van Kerkwijk M.~H.,    Kulkarni S.~R.,  2000, Nat, 408, 689

\bibitem[\protect\citeauthoryear{Hulleman, van Kerkwijk \& Kulkarni}{Hulleman
  et~al.}{2004}]{hulleman04}
Hulleman F.,  van Kerkwijk M.~H.,    Kulkarni S.~R.,  2004, A\&A, 416, 1037

\bibitem[\protect\citeauthoryear{Israel, Mereghetti \& Stella}{Israel
  et~al.}{2002}]{israel02}
Israel G.~L.,  Mereghetti S.,    Stella L.,  2002, Mem.~Soc.~Astron.~Ital., 73,
  465

\bibitem[\protect\citeauthoryear{Kaspi, Gavriil, Woods, Jensen, Roberts \&
  Chakrabarty}{Kaspi et~al.}{2003}]{kaspi03}
Kaspi V.~M.,  Gavriil F.~P.,  Woods P.~M.,  Jensen J.~B.,  Roberts M. S.~E.,
  Chakrabarty D.,  2003, ApJ, 588, L93

\bibitem[\protect\citeauthoryear{Kern \& Martin}{Kern \&
  Martin}{2002}]{kern02b}
Kern B.,  Martin C.,  2002, Nat, 417, 527

\bibitem[\protect\citeauthoryear{Lyne, Jordan \& Roberts}{Lyne
  et~al.}{2005}]{lyne05}
Lyne A.~G.,  Jordan C.~A.,    Roberts M.~E.,  2005, Monthly ephemeris, Jodrell
  Bank Crab Pulsar Timing Results.
Jodrell Bank Observatory, University of Manchester

\bibitem[\protect\citeauthoryear{{Mereghetti}, {Chiarlone}, {Israel} \&
  {Stella}}{{Mereghetti} et~al.}{2002}]{mereghetti02}
{Mereghetti} S.,  {Chiarlone} L.,  {Israel} G.~L.,    {Stella} L.,  2002, in
  Becker W.,  Lesch H.,   Tr{\" u}mper J.,  eds, Neutron Stars, Pulsars, and
  Supernova Remnants MPE Report 278, p.~29

\bibitem[\protect\citeauthoryear{Naylor}{Naylor}{1998}]{naylor98}
Naylor T.,  1998, MNRAS, 296, 339

\bibitem[\protect\citeauthoryear{Patel, Kouveliotou, Woods, Tennant, Weisskopf,
  Finger, Wilson, G{\"o}{\u g}{\"u}{\c s}, van~der Klis \& Belloni}{Patel
  et~al.}{2003}]{patel03}
Patel S.~K.,  Kouveliotou C.,  Woods P.~M.,  Tennant A.~F.,  Weisskopf M.~C.,
  Finger M.~H.,  Wilson C.~A.,  G{\"o}{\u g}{\"u}{\c s} E.,  van~der Klis M.,
   Belloni T.,  2003, ApJ, 587, 367

\bibitem[\protect\citeauthoryear{{Perna}, {Hernquist} \& {Narayan}}{{Perna}
  et~al.}{2000}]{perna00}
{Perna} R.,  {Hernquist} L.,    {Narayan} R.,  2000, ApJ, 541, 344

\bibitem[\protect\citeauthoryear{Press \& Rybicki}{Press \&
  Rybicki}{1989}]{press89}
Press W.~H.,  Rybicki G.~B.,  1989, ApJ, 338, 277

\bibitem[\protect\citeauthoryear{Smith, Tucker, Kent, Richmond, Fukugita,
  Ichikawa, Ichikawa, Jorgensen, Uomoto, Gunn, Hamabe, Watanabe, Tolea, Henden,
  Annis, Pier, McKay, Brinkmann, Chen, Holtzman, Shimasaku \& York}{Smith
  et~al.}{2002}]{smith02}
Smith J.~A.,  Tucker D.~L.,  Kent S.,  Richmond M.~W.,  Fukugita M.,  Ichikawa
  T.,  Ichikawa S.,  Jorgensen A.~M.,  Uomoto A.,  Gunn J.~E.,  Hamabe M.,
  Watanabe M.,  Tolea A.,  Henden A.,  Annis J.,  Pier J.~R.,  McKay T.~A.,
  Brinkmann J.,  Chen B.,  Holtzman J.,  Shimasaku K.,    York D.~G.,  2002,
  AJ, 123, 2121

\bibitem[\protect\citeauthoryear{Stevenson}{Stevenson}{2004}]{stevenson04}
Stevenson M.~J.,  2004, PhD thesis, University of Sheffield

\bibitem[\protect\citeauthoryear{Thompson \& Duncan}{Thompson \&
  Duncan}{1996}]{thompson96}
Thompson C.,  Duncan R.~C.,  1996, ApJ, 473, 322

\bibitem[\protect\citeauthoryear{Thompson, Lyutikov \& Kulkarni}{Thompson
  et~al.}{2002}]{thompson02}
Thompson C.,  Lyutikov M.,    Kulkarni S.~R.,  2002, ApJ, 574, 332

\bibitem[\protect\citeauthoryear{Woods \& Thompson}{Woods \&
  Thompson}{2004}]{woods04}
Woods P.~M.,  Thompson C.,  2004, in Lewin W. H.~G.,  van~der Klis M.,  eds,
  Compact Stellar X-ray Sources. CUP, Cambridge, in press (astro-ph/0406133)

\end{thebibliography}

\label{lastpage}

\end{document}